\documentclass[]{spie}

\usepackage{amsmath,amsfonts,amssymb}
\usepackage{graphicx}
\usepackage[colorlinks=true, allcolors=blue]{hyperref}
\usepackage{multicol}
\usepackage[bottom=0.6in,left=0.6in,right=0.6in]{geometry}

\def\PRL{\textit{Phys. Rev. Lett.}}
\def\PRD{\textit{Phys. Rev.} D}
\def\AA{\textit{Astron. Astrophys.}}
\def\APJ{\textit{Astrophys. J.}}
\def\SPIE{\textit{Proc. SPIE Int. Soc. Opt. Eng.}}

\newcommand{\Planck}{\textit{Planck}}
\newcommand{\Keck}{\textit{Keck}}
\newcommand{\KeckArray}{\textit{Keck Array}}
\newcommand{\BICEP}{\textsc{Bicep}}
\newcommand{\BICEPArray}{{\textsc{Bicep} Array}}

\title{The Latest Constraints on Inflationary B-modes from the BICEP/Keck Telescopes}

\author[a]{The \BICEP/\Keck\ Collaboration: P.~A.~R.~Ade}
\author[b]{Z.~Ahmed}
\author[c]{M.~Amiri}
\author[d]{D.~Barkats}
\author[e]{R.~Basu~Thakur}
\author[f]{C.~A.~Bischoff}
\author[b,g]{D.~Beck}
\author[e,h]{J.~J.~Bock}
\author[d]{H.~Boenish}
\author[i]{E.~Bullock}
\author[j]{V.~Buza}
\author[i]{J.~R.~Cheshire~IV}
\author[d]{J.~Connors}
\author[d]{J.~Cornelison}
\author[k]{M.~Crumrine}
\author[g,b]{A.~Cukierman}
\author[l]{E.~V.~Denison}
\author[d]{M.~Dierickx}
\author[m]{L.~Duband}
\author[d]{M.~Eiben}
\author[c]{S.~Fatigoni}
\author[n,o]{J.~P.~Filippini}
\author[k]{S.~Fliescher}
\author[f]{C.~Giannakopoulos}
\author[g]{N.~Goeckner-Wald}
\author[d]{D.~C.~Goldfinger}
\author[g]{J.~Grayson}
\author[d]{P.~Grimes}
\author[k]{G.~Hall}
\author[g]{G.~Halal}
\author[c]{M.~Halpern}
\author[f]{E.~Hand}
\author[d]{S.~Harrison}
\author[b]{S.~Henderson}
\author[e,h]{S.~R.~Hildebrandt}
\author[l]{G.~C.~Hilton}
\author[l]{J.~Hubmayr}
\author[e]{H.~Hui}
\author[g,b,l]{K.~D.~Irwin}
\author[g,e]{J.~Kang}
\author[d,j]{K.~S.~Karkare}
\author[g]{E.~Karpel}
\author[e]{S.~Kefeli}
\author[g]{S.~A.~Kernasovskiy}
\author[d,p]{J.~M.~Kovac}
\author[g,b]{C.~L.~Kuo}
\author[k,*]{K.~Lau}
\author[j]{E.~M.~Leitch}
\author[n]{A.~Lennox}
\author[h]{K.~G.~Megerian}
\author[e]{L.~Minutolo}
\author[e]{L.~Moncelsi}
\author[g]{Y.~Nakato}
\author[q]{T.~Namikawa}
\author[h]{H.~T.~Nguyen}
\author[e,h]{R.~O'Brient}
\author[g,b]{R.~W.~Ogburn~IV}
\author[f]{S.~Palladino}
\author[d]{M.~Petroff}
\author[m]{T.~Prouve}
\author[k,i]{C.~Pryke}
\author[d,r]{B.~Racine}
\author[l]{C.~D.~Reintsema}
\author[d]{S.~Richter}
\author[e]{A.~Schillaci}
\author[k]{R.~Schwarz}
\author[d]{B.~L.~Schmitt}
\author[s]{C.~D.~Sheehy}
\author[i]{B.~Singari}
\author[e]{A.~Soliman}
\author[d,p]{T.~St.~Germaine}
\author[e]{B.~Steinbach}
\author[a]{R.~V.~Sudiwala}
\author[e]{G.~P.~Teply}
\author[g,b]{K.~L.~Thompson}
\author[g]{J.~E.~Tolan}
\author[a]{C.~Tucker}
\author[h]{A.~D.~Turner}
\author[f,n]{C.~Umilt\`{a}}
\author[d]{C.~Verg\`{e}s}
\author[t,j]{A.~G.~Vieregg}
\author[e]{A.~Wandui}
\author[h]{A.~C.~Weber}
\author[c]{D.~V.~Wiebe}
\author[k]{J.~Willmert}
\author[d,p]{C.~L.~Wong}
\author[b]{W.~L.~K.~Wu}
\author[g]{H.~Yang}
\author[g,b]{K.~W.~Yoon}
\author[g,b]{E.~Young}
\author[g]{C.~Yu}
\author[d]{L.~Zeng}
\author[e]{C.~Zhang}
\author[e]{S.~Zhang}

\affil[a]{School of Physics and Astronomy, Cardiff University, Cardiff, CF24 3AA, United Kingdom}
\affil[b]{Kavli Institute for Particle Astrophysics and Cosmology, SLAC National Accelerator Laboratory, 2575 Sand Hill Rd, Menlo Park, CA 94025, USA}
\affil[c]{Department of Physics and Astronomy, University of British Columbia, Vancouver, British Columbia, V6T 1Z1, Canada}
\affil[d]{Center for Astrophysics, Harvard \& Smithsonian, Cambridge, MA 02138, USA}
\affil[e]{Department of Physics, California Institute of Technology, Pasadena, CA 91125, USA}
\affil[f]{Department of Physics, University of Cincinnati, Cincinnati, OH 45221, USA}
\affil[g]{Department of Physics, Stanford University, Stanford, CA 94305, USA}
\affil[h]{Jet Propulsion Laboratory, Pasadena, CA 91109, USA}
\affil[i]{Minnesota Institute for Astrophysics, University of Minnesota, Minneapolis, MN 55455, USA}
\affil[j]{Kavli Institute for Cosmological Physics, University of Chicago, Chicago, IL 60637, USA}
\affil[k]{School of Physics and Astronomy, University of Minnesota, Minneapolis, MN 55455, USA}
\affil[l]{National Institute of Standards and Technology, Boulder, CO 80305, USA}
\affil[m]{Service des Basses Temp\'{e}ratures, Commissariat \`{a} l'Energie Atomique, 38054 Grenoble, France}
\affil[n]{Department of Physics, University of Illinois at Urbana-Champaign, Urbana, IL 61801, USA}
\affil[o]{Department of Astronomy, University of Illinois at Urbana-Champaign, Urbana, IL 61801, USA}
\affil[p]{Department of Physics, Harvard University, Cambridge, MA 02138, USA}
\affil[q]{Kavli Institute for the Physics and Mathematics of the Universe (WPI), UTIAS, The~University~of~Tokyo, Kashiwa, Chiba 277-8583, Japan}
\affil[r]{Aix-Marseille  Universit\'{e},  CNRS/IN2P3,  CPPM,  13288 Marseille,  France}
\affil[s]{Physics Department, Brookhaven National Laboratory, Upton, NY 11973, USA}
\affil[t]{Department of Physics, Enrico Fermi Institute, University of Chicago, Chicago, IL 60637, USA}

\authorinfo{*Corresponding author: K. Lau, \href{mailto:kennylau@umn.edu}{kennylau@umn.edu}}
\begin{document}
\maketitle

\begin{abstract}
For the past decade, the \BICEP/\Keck\ collaboration has been operating a series of telescopes at the Amundsen-Scott South Pole Station measuring degree-scale $B$-mode polarization imprinted in the Cosmic Microwave Background (CMB) by primordial gravitational waves (PGWs). These telescopes are compact refracting polarimeters mapping about 2\% of the sky, observing at a broad range of frequencies to account for the polarized foreground from Galactic synchrotron and thermal dust emission. Our latest publication ``BK18'' utilizes the data collected up to the 2018 observing season, in conjunction with the publicly available WMAP and \Planck\ data, to constrain the tensor-to-scalar ratio $r$. It particularly includes (1) the 3-year \BICEP3 data which is the current deepest CMB polarization map at the foreground-minimum 95~GHz; and (2) the \Keck\ 220~GHz map with a higher signal-to-noise ratio on the dust foreground than the \Planck\ 353~GHz map. We fit the auto- and cross-spectra of these maps to a multicomponent likelihood model ($\Lambda$CDM+dust+synchrotron+noise+$r$) and find it to be an adequate description of the data at the current noise level. The likelihood analysis yields $\sigma(r)=0.009$. The inference of $r$ from our baseline model is tightened to $r_{0.05}=0.014^{+0.010}_{-0.011}$ and $r_{0.05}<0.036$ at 95\% confidence, meaning that the \BICEP/\Keck\ $B$-mode data is the most powerful existing dataset for the constraint of PGWs. The up-coming \BICEPArray\ telescope is projected to reach $\sigma(r) \lesssim 0.003$ using data up to 2027.
\end{abstract}

\section{Introduction}

The inflation paradigm is one of the leading candidates for a theory that describes the evolution of the early Universe. It suggests that at the time of $\approx 10^{-36}$~s after the Big Bang, the Universe underwent a brief period of exponential expansion in which its scale was increased by 50 to 60 e-folds. This scenario offers natural solutions to the horizon, flatness and magnetic monopole problems arising from the hot Big Bang expansion model~\cite{aguth}. Moreover, inflation predicts the generation of scalar perturbations in the early Universe. These perturbations are supposed to be adiabatic, nearly Gaussian and close to scale-invariant, which are consistent with high precision full-sky Cosmic Microwave Background (CMB) measurements~\cite{p2018a}.

Nevertheless, generic inflationary models make a prediction that has not yet been observed --- the existence of a background of primordial gravitational waves (PGWs). The magnitude of these PGWs can be parameterized by the tensor-to-scalar ratio $r$, which was constrained to be at least an order of magnitude smaller than one by CMB temperature data in early 2010s~\cite{p2013}. PGWs would also leave potentially detectable ``primordial $B$-modes"~\cite{seljak97,kamionkowski97} which are characteristically different from and substantially weaker than the $E$-modes in CMB polarization. Thus, the search for $B$-mode polarization can serve as a probe of PGWs. The constraints on $r$ from $B$-mode measurements, when combined with the constraint on spectral index $n_s$, can discriminate among inflation models~\cite{s4book}.

Galactic foregrounds are one of the factors obscuring the detection of primordial $B$-modes. In 2014, the publication of \BICEP2 analysis~\cite{bki} and \Planck\ 353~GHz data~\cite{pipxxx,b2k} revealed that polarized thermal emission from dust within our galaxy dominates the $B$-mode signals at high microwave frequencies. Other studies~\cite{spass} also found Galactic polarized synchrotron emission at low frequency. These facts imply that multiple-frequency measurements are necessary to account for the contribution of foregrounds. 

Gravitational lensing is another major source of $B$-modes. While polarized CMB photons were propagating from the last-scattering surface, they were deflected by gravitational lensing, and this transforms part of the $E$-modes into ``lensing $B$-modes"~\cite{sptbmode}. Therefore, in the absence of Galactic foreground, the total $B$-mode power spectrum is the combination of primordial $B$-modes and lensing $B$-modes.  As the latter component is prominent at small scales, measuring the degree-scale $B$-mode polarization emerges as one of the most promising methods for a first detection. 

The \BICEP/\Keck\ collaboration has been operating a series of small-aperture telescopes covering a broad range of frequencies for $B$-mode polarization measurement for the past decade. This proceeding discusses the latest constraints on PGWs by the \BICEP/\Keck\ data as well as the prospect of the \BICEPArray\ system. In section \ref{sec:bkseries}, we provide an overview of the \BICEP/\Keck\ program; in section \ref{sec:bk18} we show the analysis results driven by \BICEP3 and \KeckArray\ data up to the 2018 observing season. This includes their maps, power spectrum and the derived constraints on $r$ and inflation models; and in section \ref{sec:ba} we briefly outline the $r$ constraints offered by \BICEPArray\ in the future.

\section{The BICEP/\Keck\ Telescope Series}
\label{sec:bkseries}

\begin{figure}[h]
\centering
\includegraphics[width=0.7\linewidth]{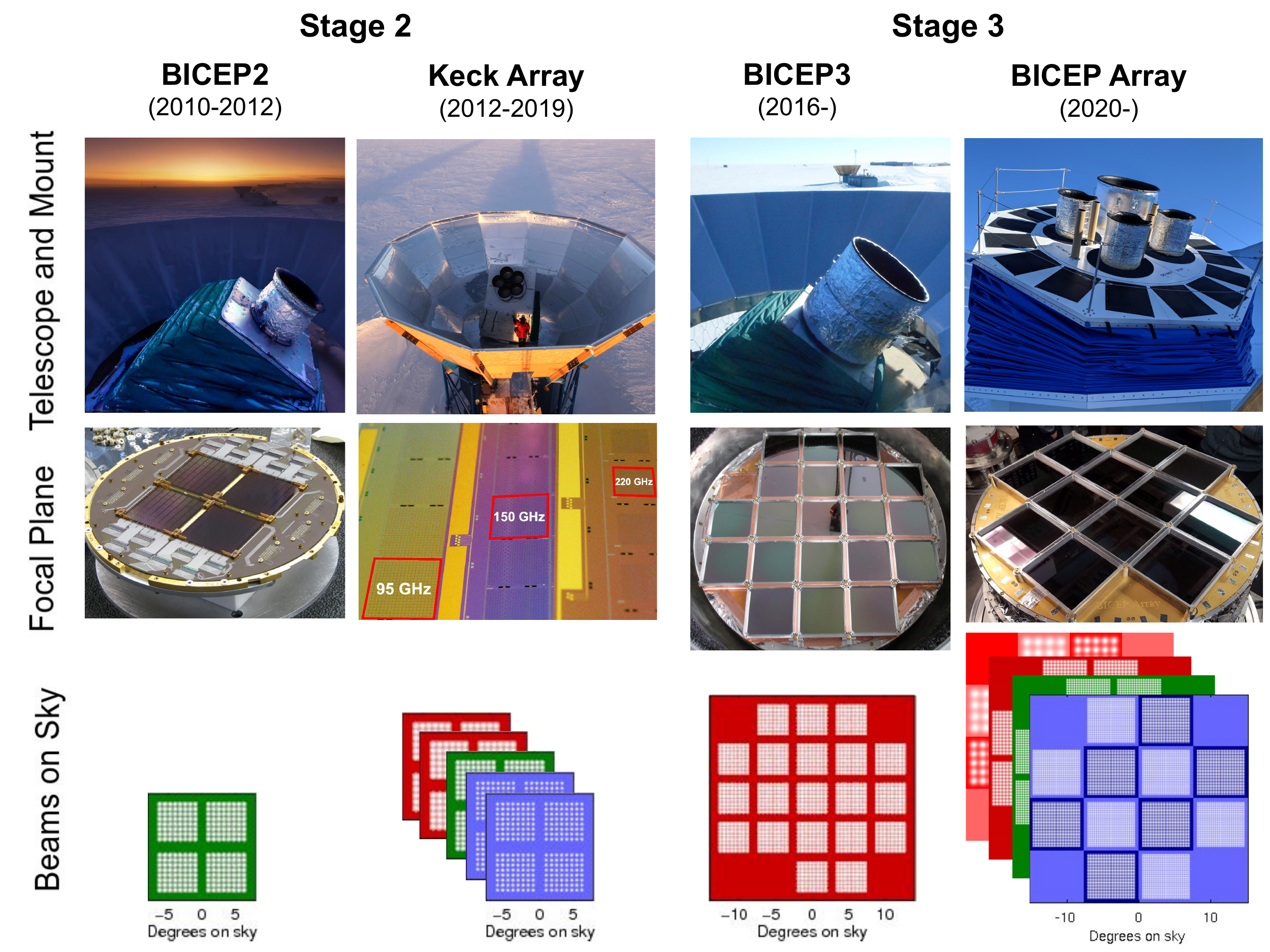}
\caption{The \BICEP/\Keck\ telescopes are Stage-2 and Stage-3 CMB ground-based experiments characterized by $\approx 10^3$ and $\approx 10^4$ detectors respectively. The top two panels show the telescopes (and their physical focal planes) operating since 2010. In the bottom panel, each plane stands for a receiver. The white dots represent the detectors and the beam size as projected on the sky, while the color of a plane denotes the observing frequency. The first receiver (30/40~GHz) of \BICEPArray\ started running in 2020, and it will be fully upgraded to have the other three receivers shown in the figure by 2023.}
\label{fig:bkseries}
\end{figure}

As illuminated in figure \ref{fig:bkseries}, the \BICEP/\Keck\ collaboration has designed, manufactured and operated four dedicated telescopes to measure CMB $B$-mode polarization since 2010. \BICEP2~\cite{bkii}, a single-receiver telescope observing at 150~GHz from 2010-2013, had the conceptual designs of detector and optic employed by subsequent upgrades. Its aperture size was~$\approx 250$~mm and its focal plane contained four $4\,'' \times 4\,''$ wafers with a total of $\approx$ 500 bolometric detectors --- the antenna-coupled transition edge sensors (TES). Each pair of co-located detectors consisted of interleaved arrays of orthogonal slot antennas so that the pair difference timestream is a measure of polarized emission from the sky. During observations, the focal plane was cooled to the temperature of $\approx$ 250~mK by a 3-stage Helium sorpotion refrigerator, and the detectors were read out by time-division multiplexing SQUID amplifiers. The telescope insert with a compact, on-axis, two-lens refractor optical design was cooled to 4~K to minimize in-band photon loading. The receiver was further coupled with a cylindrical co-moving forebaffle and surrounded by a reflective ground shield for optical shielding. 

The \KeckArray\ telescope~\cite{bkiv} incorporated five \BICEP2-style receivers with pulse tube cooled cryostats. All receivers, each with $\approx$ 500 detectors, initially observed at 150~GHz. As the focus of CMB experiments targeting primordial $B$-modes shifted from raw sensitivity improvement to multiple-frequency observation, it gradually switched to cover 95~GHz and 220~GHz from 2014 to 2019 to constrain the contribution from dust foreground emission.

\BICEP3~\cite{bkxv} has been observing at the foreground-minimum 95~GHz since 2016. It is a single-receiver telescope with doubled aperture size $\approx$ 500~mm for a larger instantaneous field of view. \BICEP3 hosts $\approx$ 2500 detectors packed into twenty $4\,'' \times 4\,''$ focal plane modules, and is equivalent to about eight of the \Keck\ 95~GHz receivers in terms of map depth.

In late 2019, \Keck\ was decommissioned and replaced by the \BICEPArray\ telescope~\cite{hui,moncelsi} which will eventually carry four \BICEP3-style receivers. It will be discussed separately in section~\ref{sec:ba}. 

All four telescopes were deployed to observe at the Amundsen-Scott South Pole Station. \BICEP2 and \BICEP3 were installed on the same mount at the Dark Sector Laboratory (DSL), while \Keck\ and \BICEPArray\ were located at the Martin A. Pomerantz Observatory (MAPO) about 200~m away. The high altitude ($\approx$ 2800~m) and extreme cold ($\approx -70 \, ^\circ\mathrm{C}$ in winter) of the South Pole suppress precipitable water vapor. The atmosphere is hence exceptionally stable and transparent at microwave frequencies, providing one of the best observation sites for ground-based CMB telescopes. Other advantages include the 6-month nighttime, the continuous visibility of the ``Southern Hole'' (i.e. a low-foreground patch of the sky) and minimal light pollution.

The \BICEP/\Keck\ observation strategy is to concentrate on one of the cleanest patches of the sky available from the South Pole. The observing field is centered at RA 0~hr, Dec. -57.5$^\circ$. With the elevation fixed, the telescope moves back and forth along the azimuth direction to scan $\approx 100^\circ$ on the sky. In 50~minutes of time, it conducts 100 ``half-scans'' in this way, and these yield a single observation unit, a ``scan set''.  After several tens of scan sets, the telescope spins along its boresight (``deck angle'' rotation) to another deck angle choice to cover 4 to 8 distinct angles for the demodulation of $Q$ and $U$ signals. Moreover, a total of 21 evenly divided elevation offsets between Dec. $-55^\circ$ to $-60^\circ$ expand the effective observation area to $\approx 400$~square~degrees for \BICEP2/\Keck, or $\approx 585$~square~degrees for \BICEP3/\BICEPArray. Over a full observation season, each telescope can accumulate $\approx$ 4500 scan sets, and around 60\% of them survive data cuts. That means abundant observation time is devoted to $\approx$ 2\% of the sky for deep maps. This is a typical strategy for a first detection. 

\section{BK18 Maps, Power spectrum and Constraints on Inflation Models}
\label{sec:bk18}

In our previous mainline publication (hereafter ``BK15")~\cite{bk15}, we utilize \BICEP2 and \Keck\ \linebreak 95/150/220~GHz data up to the 2015 observing season, in conjunction with external data, to yield the constraint $r_{0.05} < 0.06$ (evaluated at the pivot scale $k=0.05 \, \textup{Mpc}^{-1}$) at 95\% confidence. Our latest publication (hereafter ``BK18")~\cite{bk18} adds three more years of data from \Keck\ and \BICEP3. While the BK15 dataset consists of 4/17/2 \Keck\ receiver-years at 95/150/220~GHz respectively, the BK18 dataset is equivalent to about 28/18/14 \Keck\ receiver-years at 95/150/220~GHz respectively. 

\begin{figure}[t]
\centering
\includegraphics[width=0.6\linewidth]{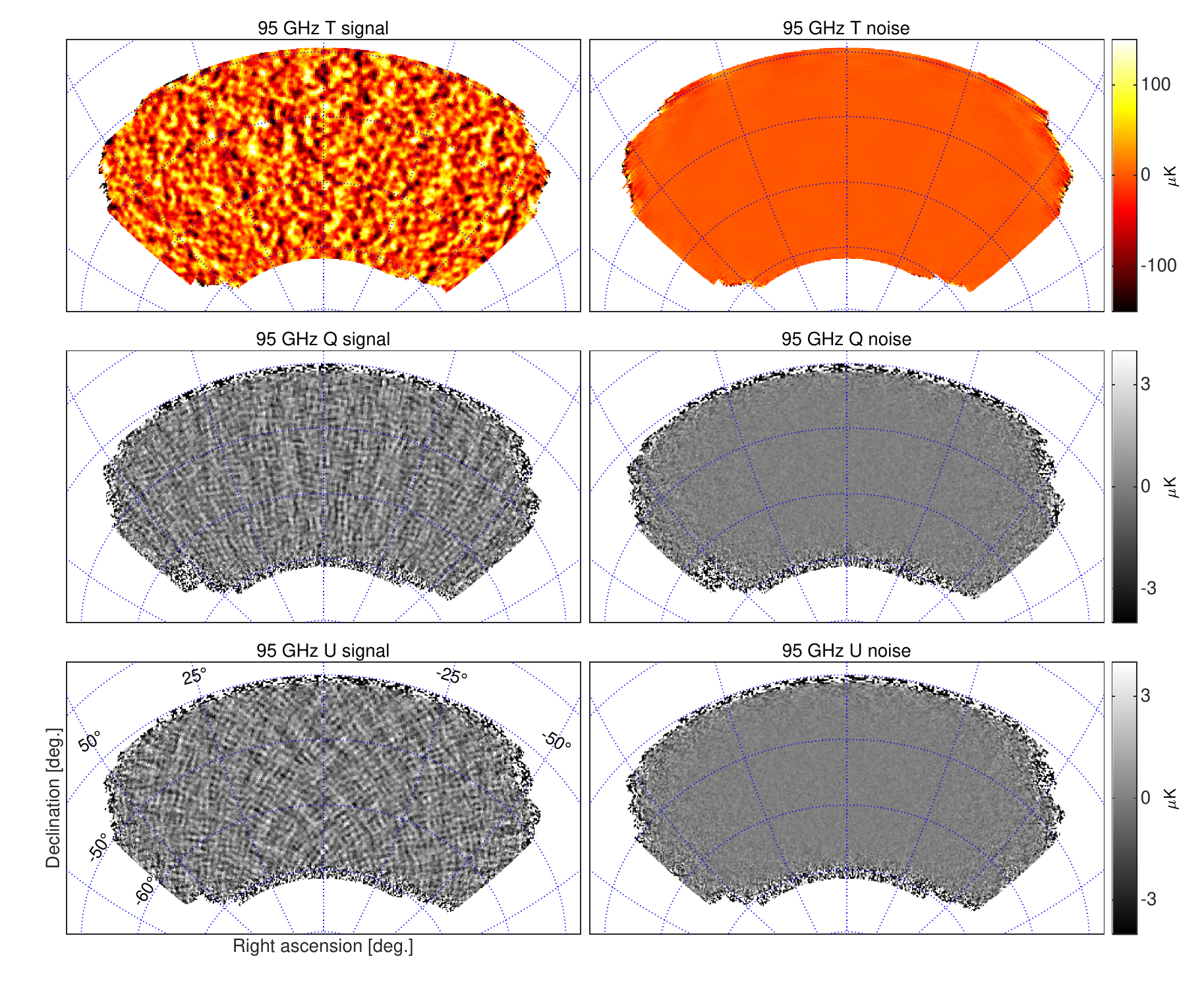}
\caption{The \BICEP3 95~GHz $T$, $Q$, $U$ maps from the BK18 paper are displayed in the left column. These signal maps are made using \BICEP3 data from the 2016-2018 observing seasons. The right column shows maps of a noise realization produced by randomly and evenly flipping the sign of scan sets during the map coadding process. Note that the signal maps in general appear different from the full-sky measurement of the same sky patch since they are heavily filtered by beam smoothing, timestream processing and deprojection.}
\label{fig:b95maps}
\end{figure}

The analysis starts by processing the detector timestreams of scan sets. They are first relatively calibrated to eliminate the gain difference between two detectors in a pair. To remove the $1/f$ atmospheric noise, ground signals and magnetic pickup, third-order polynomial filtering and scan-synchronous subtraction are applied to the timestreams of each half-scan. They are also run through the deprojection process for the removal of $T \rightarrow P$ leakage, and two distinct rounds of data cut are further applied to remove data from bad weather periods. Pair difference timestreams from multiple deck angle measurements can then be demodulated into $Q$ and $U$ signals via a rotation matrix inversion. Finally, they can be binned into $T$, $Q$, $U$ maps with the pointing trajectory information. 

\begin{figure}[t]
\centering
\includegraphics[width=0.6\linewidth]{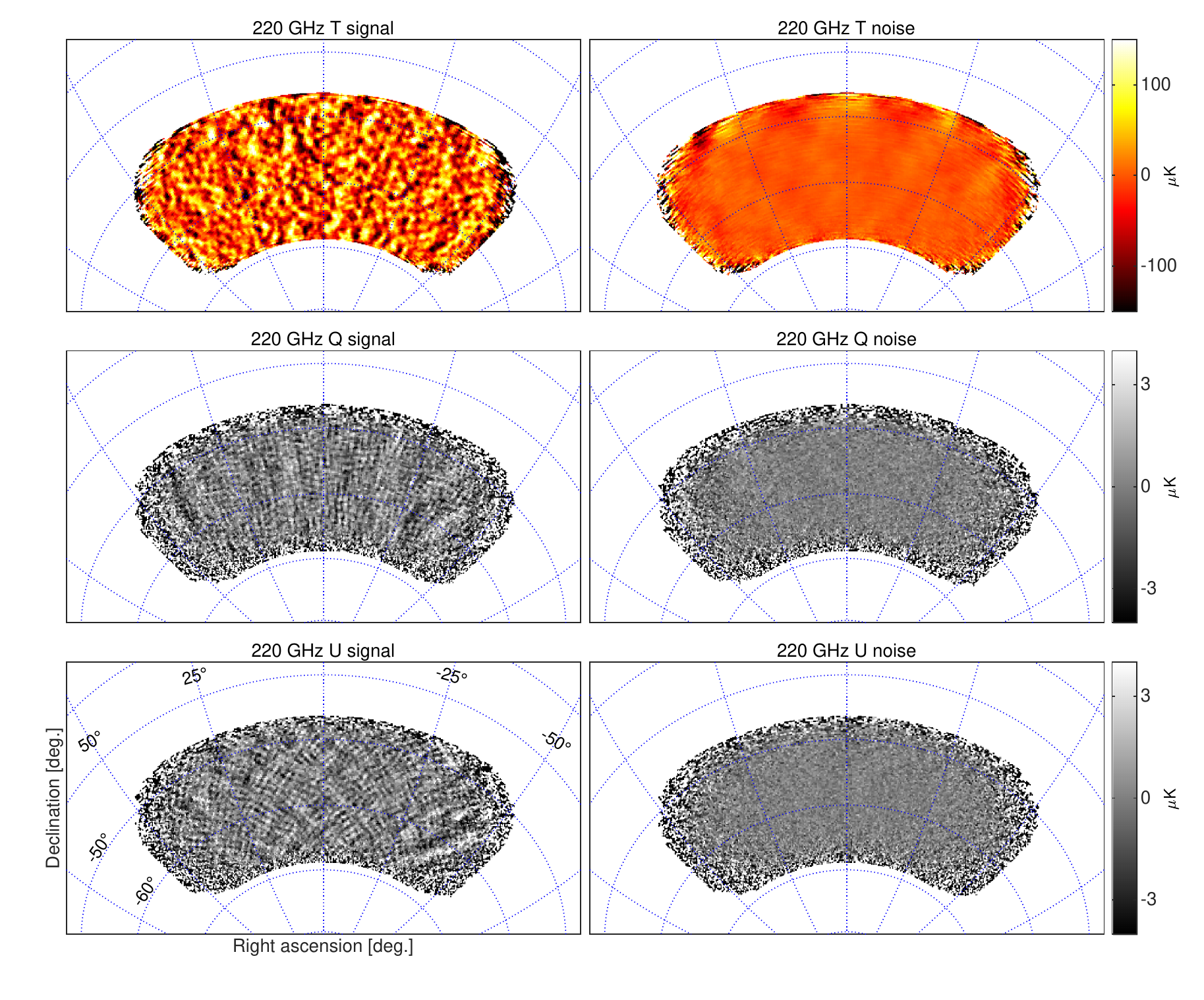}
\caption{\Keck\ 220~GHz $T$, $Q$, $U$ regular and noise maps from the BK18 paper. They are made using data equivalent to 14 \Keck\ receiver-years. Note that the irregularities around the lower right corners of the $Q$ and $U$ maps are due to polarized dust emission.}
\label{fig:k220maps}
\end{figure}

Figure \ref{fig:b95maps} and figure \ref{fig:k220maps} show the \BICEP3 95~GHz and \Keck\ 220~GHz $T$, $Q$, $U$ maps published in the BK18 paper. They contain the \BICEP3 and \Keck\ data up to the 2018 observing season. The comparison between regular maps and noise maps indicates high S/N detection of polarization signals. With the expanded field of view, and an order of magnitude increase in detector number, the \BICEP3 $Q$/$U$ maps, which are the current deepest 95~GHz CMB polarization maps, achieve a map depth of 2.8~$\mu$K-arcmin. The \Keck\ 220~GHz $Q$/$U$ map depth of 8.8~$\mu$K-arcmin also far exceeds the sensitivity provided by \Planck\ 353~GHz measurement in the same sky patch. 

Apart from the modest smoothing due to the change of beam size, these two sets of maps display strong correlation over frequency and show clear ``+" structure in $Q$ and ``$\times$" structure in $U$, implying that they detect the dominating $E$-mode signals from the $\Lambda$CDM model. The exception is the small ``anomalies" around the lower right corners of the \Keck\ polarization maps. This pattern is contributed by the polarized thermal dust emission. It obeys the expected spectral behavior that it is strong at 220~GHz while it becomes nearly invisible at 95~GHz.

In order to handle the contribution from the $\Lambda$CDM model, the foreground emission and $r$, we use the same multi-component multi-spectral likelihood analysis as in BK15. In our baseline model, we first describe the $BB$ foreground cross spectrum between maps with frequency $\nu_1$ and $\nu_2$ by
\begin{equation}
  \mathcal{D}_{\ell,BB}^{\nu_1 \times \nu_2} =
  A_\mathrm{d} f_\mathrm{d}^{\nu_1} f_\mathrm{d}^{\nu_2} \left( \frac{\ell}{80} \right)^{\alpha_\mathrm{d}} +
  A_\mathrm{s} f_\mathrm{s}^{\nu_1} f_\mathrm{s}^{\nu_2} \left( \frac{\ell}{80} \right)^{\alpha_\mathrm{s}} +
  \epsilon \sqrt{A_\mathrm{d}A_\mathrm{s}} \, (f_\mathrm{d}^{\nu_1} f_\mathrm{s}^{\nu_2} + f_\mathrm{s}^{\nu_1} f_\mathrm{d}^{\nu_2}) \left( \frac{\ell}{80} \right)^{(\alpha_\mathrm{d} + \alpha_\mathrm{s}) / 2}, 
  \label{eq:BB_model}
\end{equation}
where $A_\mathrm{d}$ specifies the dust power amplitude at pivot frequency $\nu = 353$~GHz and $\ell = 80$; $A_\mathrm{s}$ is similarly the synchrotron power amplitude at $\nu = 23$~GHz; $\epsilon$ is the correlation parameter between dust and synchrotron; $\alpha_\mathrm{d}$ and $\alpha_\mathrm{s}$ are angular power spectrum power law indexes; and $f_\mathrm{d}$ and $f_\mathrm{s}$ are frequency scaling accounting for the spectral energy density (SED) of foreground integrated in the actual bandpasses of a given frequency. The synchrotron SED is modeled by a power law in frequency with index $\beta_\mathrm{s}$ while the dust SED is modeled by the modified blackbody with $T_\mathrm{d} = 19.6 \,$K and frequency power law index $\beta_\mathrm{d}$. We also fix the scalar $\Lambda$CDM parameters to be the \Planck\ 2018 best fit values. Therefore, we have the $\Lambda$CDM+dust+synchrotron+$r$ baseline model with 8 parameters: $A_\mathrm{d}$, $A_\mathrm{s}$, $\alpha_\mathrm{d}$, $\alpha_\mathrm{s}$, $\beta_\mathrm{d}$, $\beta_\mathrm{s}$, $\epsilon$ and $r$.

On the data side, we have $BB$ power spectra. Four internal maps (\BICEP3 95~GHz, \Keck\ 95~GHz, \BICEP2/\Keck\ 150~GHz and \Keck\ 220~GHz) are used with multiple reprocessed WMAP (23 \& 33~GHz) and $Planck$ (30, 44, 143, 217 \& 353~GHz) maps in our observation field to increase the total maps involved to 11. These maps are then purified and inverse noise apodized~\cite{matrix} to yield 66 $BB$ auto- and cross-spectra. 

\begin{figure}[t]
\centering
\includegraphics[width=0.67\linewidth]{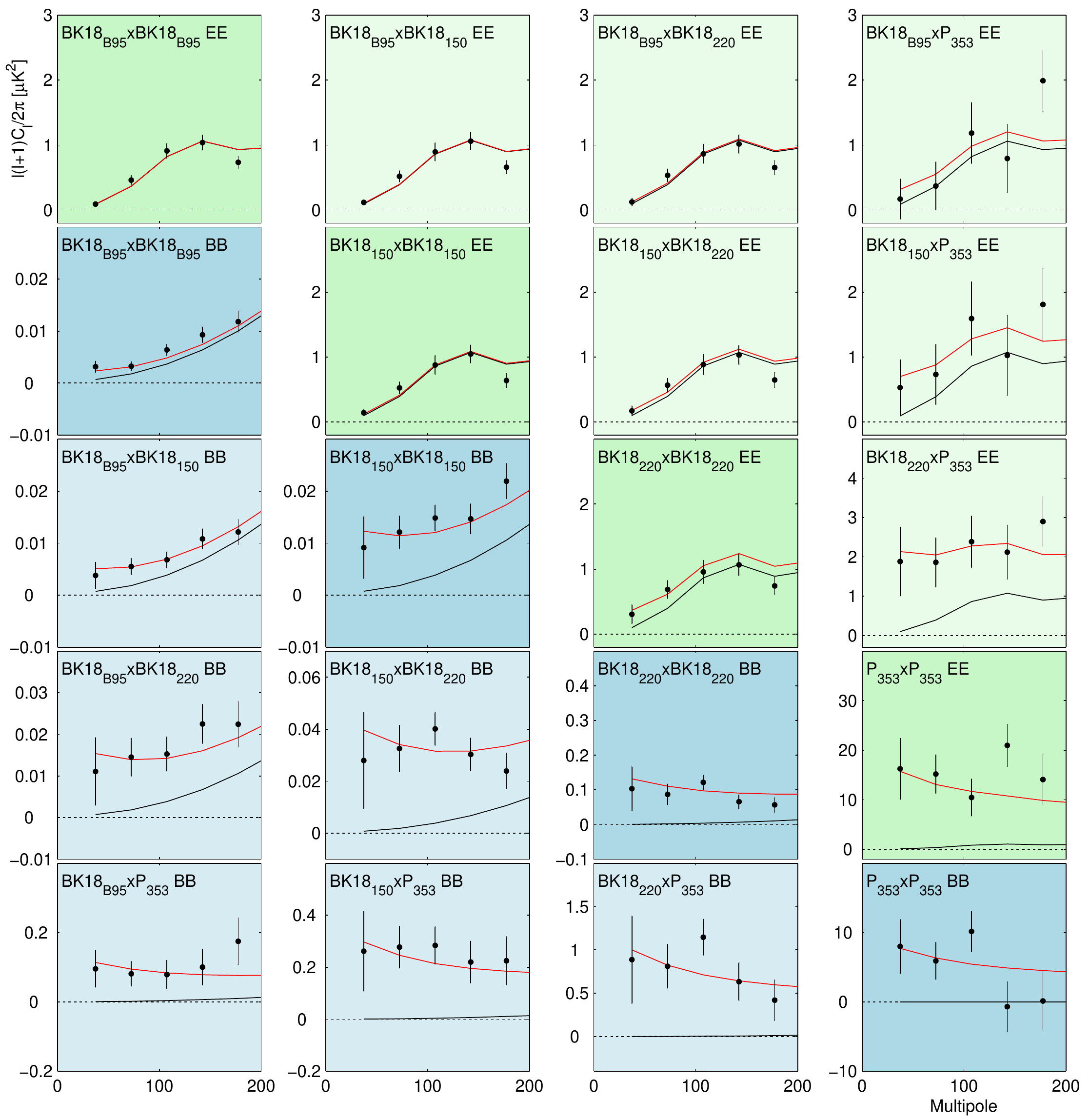}
\caption{$BB$ (blue) and $EE$ (green) auto- and cross-spectra from \BICEP3 95~GHz, \BICEP2/\Keck\ 150~GHz, \Keck\ 220~GHz and \Planck\ 353~GHz maps. The black lines are the $\Lambda$CDM model expectation values, while the red lines are the $\Lambda$CDM+foreground expectation values from the foreground best fit of our previous BK15 analysis. The $EE$ spectra are computed as a demonstration under the assumption $EE$/$BB$ = 2 for dust. $EE$ spectra are not included in our likelihood analysis.}
\label{fig:bk18spectrum}
\end{figure}

Figure \ref{fig:bk18spectrum} shows a subset of $BB$ and $EE$ spectra as a demonstration. In particular, all \BICEP3 auto- and cross-spectral bandpowers are consistent with the $\Lambda$CDM+foreground model lines which use the foreground best fit from our previous BK15 analysis with \textit{no} \BICEP3 data! This agreement is hence a validation of our foreground model at the current noise level since it correctly predicted weak foreground emission at 95~GHz. The small error bars of the \BICEP3 $BB$ bandpowers further suggest these data should provide strong constraining power on $r$.

\begin{figure}[t]
\begin{minipage}{0.60\linewidth}
\centerline{\includegraphics[width=1\linewidth]{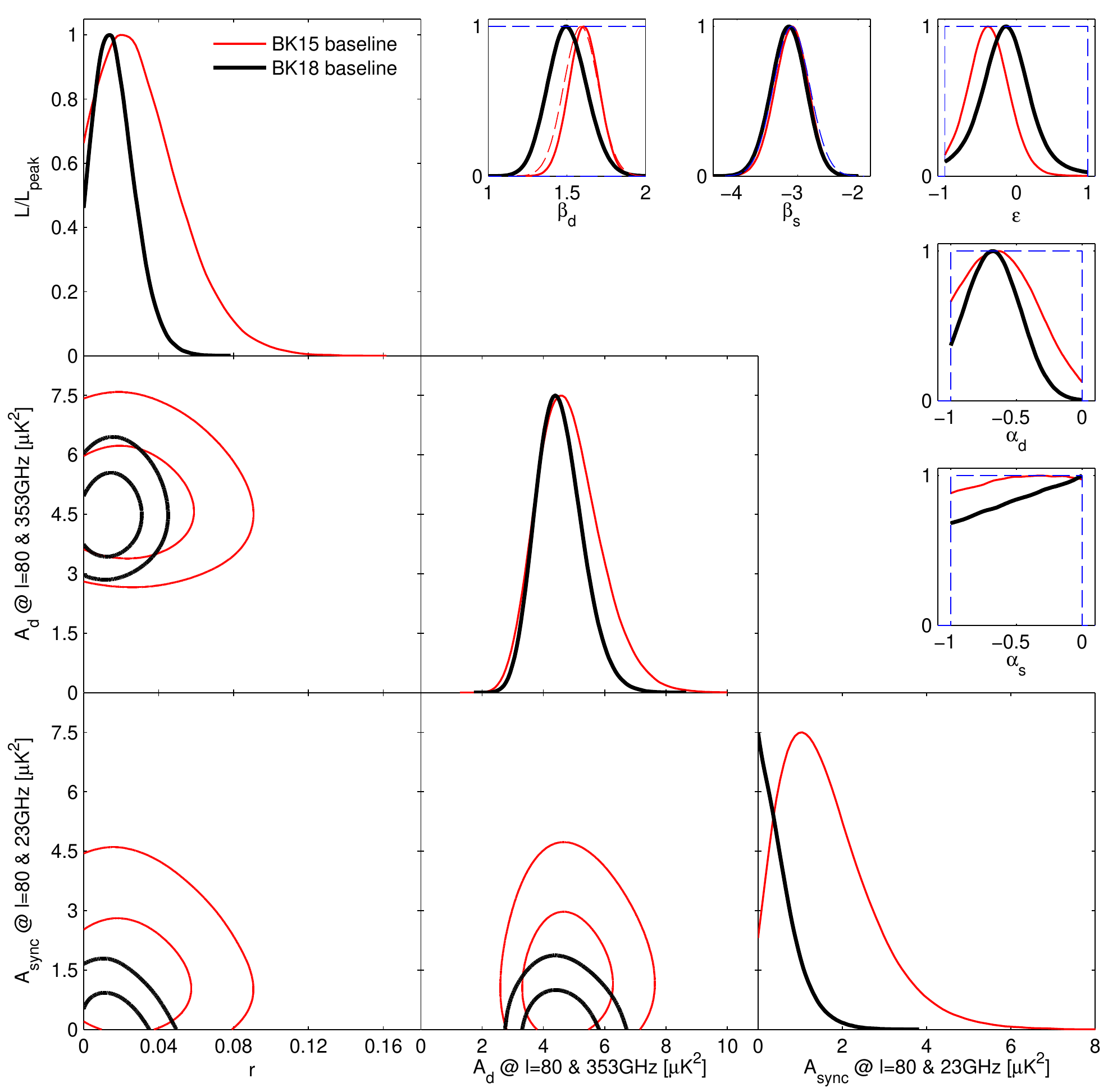}}
\end{minipage}
\begin{minipage}{0.40\linewidth}
\centerline{\includegraphics[width=1\linewidth]{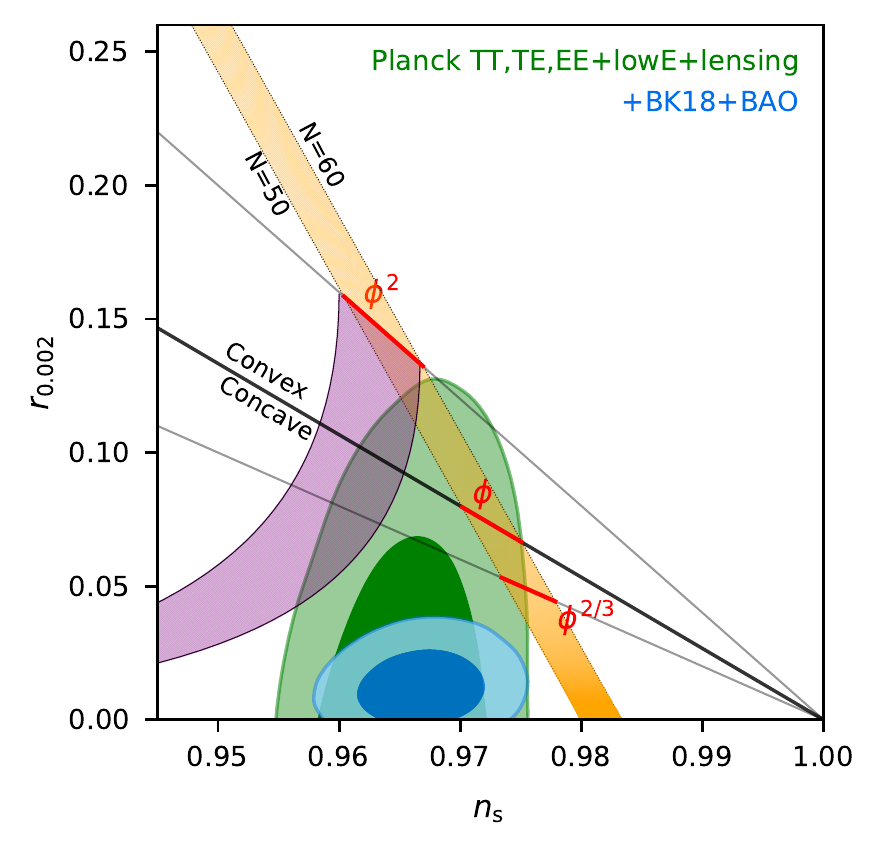}}
\end{minipage}
\caption{\textit{Left}: \texttt{CosmoMC} likelihood results for the \BICEP/\Keck\ baseline model. Selected 1D and 2D marginalized posteriors are shown. The red faint curves are the results from BK15 while the black solid curves are the results of BK18. The dashed blue and red lines show priors on foreground parameters. The analysis method is the same as in BK15, except the $\beta_\mathrm{d}$ prior based on \Planck\ data from other regions of the sky is removed this time due to the improved sensitivity of BK18. \textit{Right}: Constraints in the $r$ vs. $n_s$ plane. The purple and orange bands are natural inflation and monomial inflation respectively. The blue contour shows the updated constraint after adding BK18 and BAO data to the \Planck\ baseline analysis. The $r$ posterior is tightened from $r_{0.05} < 0.11$ to $r_{0.05} < 0.035$ at 95\% confidence.}
\label{fig:cosmomc}
\end{figure}

We then supply our real $BB$ power spectra, bandpower covariance matrix, bandpower window function and baseline model to \texttt{CosmoMC}~\cite{cosmomc} to evaluate the Hamimeche-Lewis (HL) likelihood~\cite{hl}. Figure \ref{fig:cosmomc} shows the results with the BK15 comparison. The constraint of $r_{0.05}$ from the baseline analysis, after marginalized over foreground parameters, evolves from $r_{0.05} = 0.020^{+0.021}_{-0.018}$ to $r_{0.05}=0.014^{+0.010}_{-0.011}$. The confidence interval of $r_{0.05}$ is halved, dropping from $r_{0.05} < 0.07$ to $r_{0.05} < 0.036$ at 95\% confidence. The constraint on dust amplitude is improved from $A_\mathrm{d} = 4.6^{+1.1}_{-0.9}$~$\mu$K$^2$ to $A_\mathrm{d} = 4.4^{+0.8}_{-0.7}$~$\mu$K$^2$ due to the additional \Keck\ 220~GHz data. 

Figure \ref{fig:cosmomc} also shows the constraints on the $r$ vs. $n_s$ plane as in figure 28 of Ref.~\cite{p2018b}. Since this analysis is instead a \Planck$+$BK18 joint fit, we vary those 8 parameters as well as other nuisance and scalar parameters including $n_s$. With the BK18 data it manifests unprecedented discrimination power on inflation models --- the natural inflation and monomial inflation models now lie outside the 95\% contour. Moreover, as adding \Planck\ temperature and other data only gives a slightly improved upper limit $r_{0.05} < 0.035$, it implies the progress on the constraint of $r$ is now entirely driven by $B$-mode measurements.

\section{The Prospect of BICEP Array}
\label{sec:ba}

The \BICEPArray\ (BA) telescope is the successor to \KeckArray. BA will have four \BICEP3-style receivers covering six distinct bands centered at 30, 40, 95, 150, 220 and 270~GHz. It will be equipped with $\approx 30,000$ detectors, representing another order of magnitude increase in detector number relative to \BICEP3. At the end of 2019, we installed a new, larger BA mount at MAPO and deployed the first BA 30/40~GHz receiver for the constraint of Galactic synchrotron emission. Since then, it has been observing with three adapted \Keck\ receivers on the same mount. It will be followed by the deployments of 95~GHz and 150~GHz receivers in 2022, and the 220/270~GHz receiver in 2023. Observations will continue to at least the end of 2027.

\begin{figure}[t]
\centering
\includegraphics[width=0.8\linewidth]{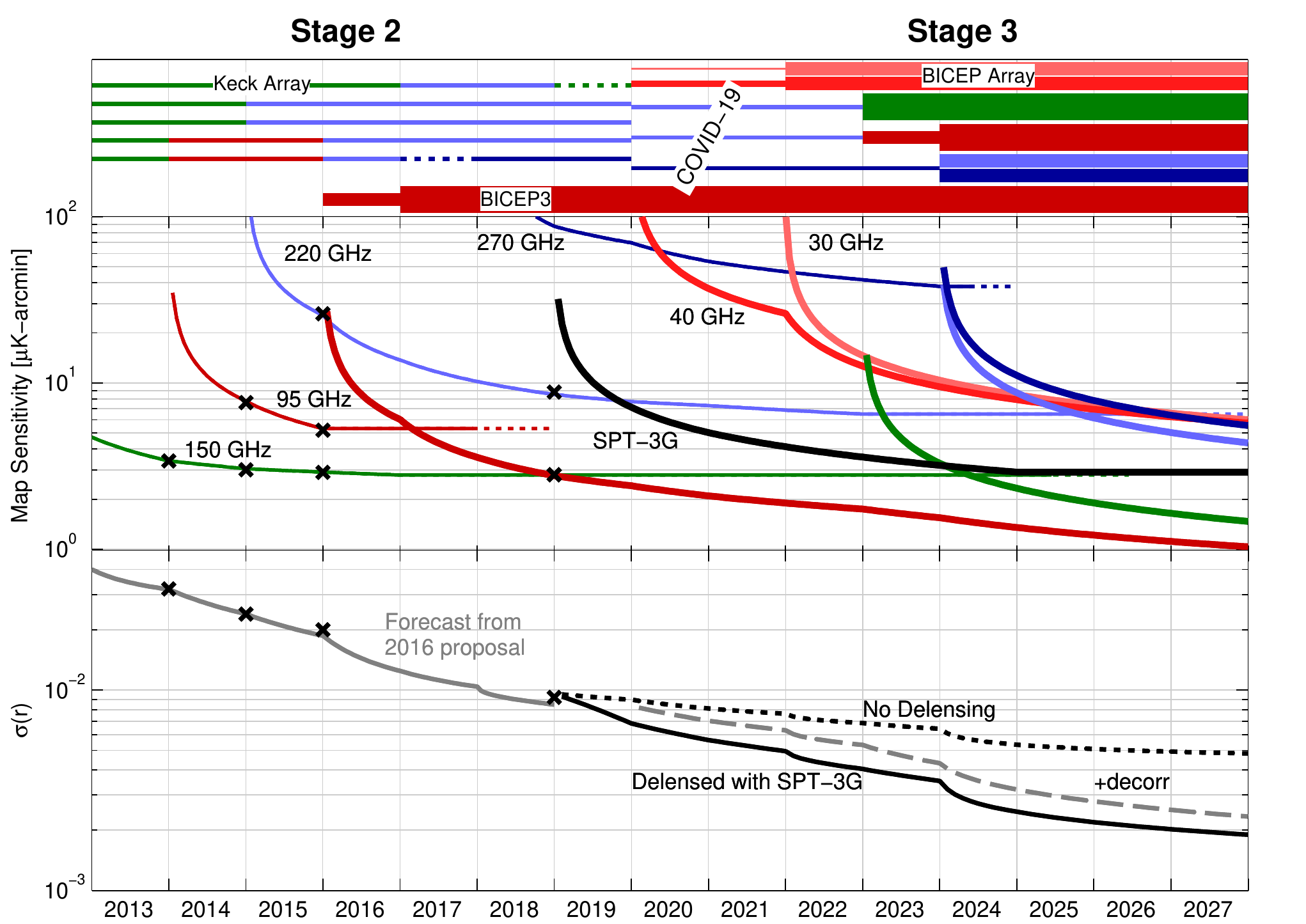}
\caption{The forecasts for the \BICEP/\Keck\ program. The top panel shows the plan of observation. The thickness of each line represents the detector number in a single receiver while its color represents the center frequency labeled in the next panel; the middle panel shows the corresponding map depth over time as lines. The cross marks are published values including those in BK15 and BK18. The sensitivity beyond 2018 is predicted by scaling based on achieved performance. The black SPT-3G line is the combined 95/150~GHz map depth projected from the achieved sensitivity in the 2019 and 2020 seasons; the bottom panel transforms map depth into $\sigma(r)$. Cross marks are again the published values. The good agreement between these cross marks and projection lines indicates that the forecasting method is realistic. Two independent forecasts (solid and dotted lines) highlight the significance of delensing. Including dust decorrelation in the delensing case (the dashed line) results in slightly higher $\sigma(r)$. }
\label{fig:bkforecast}
\end{figure}

The observation plan and sensitivity forecast for \BICEPArray\ is presented in figure \ref{fig:bkforecast}. The post-BK18 progress on $\sigma(r)$ will soon stall as the constraint will become limited by lensing variation. We therefore have been working with the South Pole Telescope (SPT) collaboration to use the overlapping SPT-3G maps for ``delensing"~\cite{bkspt}. Following this strategy, even after 2 years of COVID delay, it is projected to give $\sigma(r) \lesssim 0.003$ using data up to 2027. This will either confirm or rule out other popular classes of inflation models.

\section*{Acknowledgments}

The collaboration would like to give special thanks to our heroic winter-overs Robert Schwarz, Steffen Richter, Sam Harrison, Grantland Hall and Hans Boenish. K. Lau again expresses his gratitude to the organizers of the Rencontres de Moriond for holding a safe and successful conference during the pandemic.

\begin{multicols}{2}

\end{multicols}

\end{document}